\documentclass[conference]{IEEEtran}
\usepackage{cite}
\usepackage{amsmath,amssymb,amsfonts}
\usepackage{algorithmic}
\usepackage{graphicx}
\usepackage{textcomp}
\usepackage{xcolor}
\usepackage{amsmath,amssymb,amsfonts}
\usepackage{textcomp}
\usepackage{graphicx}
\usepackage{times}
\usepackage{microtype}
\usepackage[UKenglish]{babel}
\usepackage{graphicx,dblfloatfix}
\usepackage{blindtext}
\usepackage{mathtools}
\usepackage{multirow}
\usepackage{tabularx}
\usepackage{subcaption}
\usepackage{hhline}
\usepackage{arydshln}
\usepackage{listings}
\usepackage{adjustbox}
\usepackage{color,soul}
\usepackage[margin=1in]{geometry}
\begin{document}

\title{FPGA or GPU? Analyzing comparative research for application-specific guidance\\

}

\author{
    \IEEEauthorblockN{Arnab A Purkayastha\IEEEauthorrefmark{1}, Jay Tharwani\IEEEauthorrefmark{2}, Shobhit Aggarwal\IEEEauthorrefmark{3}}
    \IEEEauthorblockA{\IEEEauthorrefmark{1}Western New England University, Member IEEE, Springfield, MA, USA\\
    Email: arnab.purkayastha@wne.edu}
    \IEEEauthorblockA{\IEEEauthorrefmark{2} Member IEEE, Charlotte, NC, USA\\
    Email: jtharwan@alumni.uncc.edu}
     \IEEEauthorblockA{\IEEEauthorrefmark{3}{Department of ECE}, {The Citadel}, Charleston, NC, USA\\
    shobit.aggarwal@citadel.edu}

}

\maketitle

\begingroup
\renewcommand\thefootnote{}\footnote{\textbf{This work has been accepted for publication in the proceedings of IEEE Southeastcon 2025. This is the author’s preprint version. The final published version is available on IEEE Xplore.}}
\endgroup


\begin{abstract}
The growing complexity of computational workloads has amplified the need for efficient and specialized hardware accelerators. Field Programmable Gate Arrays (FPGAs) and Graphics Processing Units (GPUs) have emerged as prominent solutions, each excelling in specific domains. Although there is substantial research comparing FPGAs and GPUs, most of the work focuses primarily on performance metrics, offering limited insight into the specific types of applications that each accelerator benefits the most. This paper aims to bridge this gap by synthesizing insights from various research articles to guide users in selecting the appropriate accelerator for domain-specific applications. By categorizing the reviewed studies and analyzing key performance metrics, this work highlights the strengths, limitations, and ideal use cases for FPGAs and GPUs. The findings offer actionable recommendations, helping researchers and practitioners navigate trade-offs in performance, energy efficiency, and programmability.

\end{abstract}
 
\vspace*{0.5 cm}

\section{Introduction}
\label{sec:Introduction}

In recent years, the demand for computational power has increased across a wide range of domains, including Artificial Intelligence (AI), High-Performance Computing (HPC), and Real-time embedded systems to name a few. This has necessitated the use of hardware accelerators capable of delivering high performance while maintaining energy efficiency. Among the most prominent accelerators are Field Programmable Gate Arrays (FPGAs) and Graphics Processing Units (GPUs), each offering unique capabilities and trade-offs.

FPGAs are known for their reconfigurability, allowing users to design custom hardware tailored to specific applications. This flexibility leads to better optimization in terms of latency and power efficiency, making FPGAs particularly well-suited for real-time, power-sensitive, and application-specific tasks. However, the steep learning curve \cite{jason, socctaxonomy} and complex development processes associated with FPGA programming present significant challenges to broader adoption.

GPUs, on the other hand, excel in general-purpose parallel processing and have become a standard choice for data-intensive tasks, especially in AI and deep learning. With robust software ecosystems and mature programming frameworks like CUDA and OpenCL, GPUs offer ease of development and rapid prototyping, making them the preferred option for tasks requiring high throughput and scalability.

While both accelerators have been extensively studied, most research has focused primarily on performance comparisons, often neglecting application-specific guidance. This leaves a critical gap for practitioners seeking to understand when and why one accelerator may be preferable over the other in a given domain. Addressing this gap requires a comprehensive review of existing studies, categorized by application type, and evaluated using a consistent set of performance metrics.

This paper aims to fill that void by synthesizing insights from numerous comparative studies and categorizing them into key application domains, including HPC, AI, and vision processing. By systematically analyzing these studies, we provide a practical framework for selecting the appropriate hardware accelerator based on the specific needs of various applications. In addition, we discuss emerging trends, current challenges, and future directions in FPGA and GPU research.

In summary, the key contributions of this work are listed as follows:-
\begin{itemize}
  \item \textbf{Domain-specific categorization}: A systematic classification of existing comparative studies into key application domains, including HPC, AI, and vision processing.
  \item \textbf{Comprehensive analysis}: An in-depth analysis of FPGAs and GPUs based on performance metrics such as throughput, latency, energy efficiency, and programmability.
  \item \textbf{Actionable guidance}: Practical recommendations and guidelines for selecting the appropriate hardware accelerator based on application-specific requirements.
  \item \textbf{Future outlook}: Identification of current research gaps, emerging trends, and potential future directions in FPGA and GPU-based acceleration.
\end{itemize}

\vspace{0.3cm}
\section{Background}
\label{sec:back}

The rapid evolution of computing technology has been driven by the increasing demands of real-world applications across various domains, such as artificial intelligence (AI), high-performance computing (HPC), and embedded systems. Traditionally, Central Processing Units (CPUs) have served as the backbone of computing systems, providing general-purpose processing capabilities. However, as the complexity of workloads has grown, the need for specialized accelerators has become apparent.

CPUs were primarily designed for sequential task execution and general-purpose workloads. While they offer flexibility and ease of programmability, their performance is limited when it comes to highly parallelizable tasks. This limitation has paved the way for the adoption of accelerators like FPGAs and GPUs in applications requiring significant computational power.

GPUs have become standard in fields like AI, deep learning, and scientific simulations due to their ability to perform massive parallel processing. They are particularly well-suited for tasks that involve large-scale data processing and matrix operations, such as training deep neural networks. The availability of mature programming environments like CUDA and OpenCL has further contributed to the widespread adoption of GPUs.

On the other hand, FPGAs offer a different approach to acceleration by allowing custom hardware design tailored to specific applications. This flexibility enables highly optimized pipelines, leading to better performance in real-time, latency-sensitive, and power-constrained environments. FPGAs are commonly used in applications such as signal processing, cryptography, and embedded vision systems. 

Despite their unique strengths, choosing the right accelerator for a given application remains a challenging task. GPUs and FPGAs differ significantly in architecture, programming models, and performance characteristics. GPUs excel in throughput and ease of development, while FPGAs offer superior energy efficiency and application-specific optimization. Without adequate knowledge of which architecture is better for a specific domain, users may struggle to fully exploit the potential of these accelerators. Table \ref{fpgagpucomp} shows a summarized comparasion between the two. 

Furthermore, domain-specific applications require an understanding of domain-specific architectures to achieve optimal performance. For instance, while GPUs may be ideal for AI model training, FPGAs might be better suited for AI inference in edge devices where power consumption and latency are critical.

Although numerous works have investigated and compared these accelerators, much of the existing research focuses on performance metrics without providing sufficient application-specific guidance. This work addresses that gap by offering a domain-specific analysis of FPGAs and GPUs, helping users make informed decisions about which accelerator to choose based on the requirements of their applications.

\begin{table*}[h]
\centering
\caption{Comparison of FPGAs and GPUs based on key performance metrics.}
\label{fpgagpucomp}
\begin{tabular}{|l|l|l|}
\hline
\multicolumn{1}{|c|}{\textbf{Criteria}} &
  \multicolumn{1}{c|}{\textbf{FPGA}} &
  \multicolumn{1}{c|}{\textbf{GPU}} \\ \hline
\textbf{Throughput} &
  \begin{tabular}[c]{@{}l@{}}Moderate, depends on custom pipeline \\ design.\end{tabular} &
  High, excels in massively parallel tasks. \\ \hline
\textbf{Latency} &
  \begin{tabular}[c]{@{}l@{}}Low, highly suitable for real-time \\ applications.\end{tabular} &
  \begin{tabular}[c]{@{}l@{}}Higher compared to FPGAs, but acceptable \\ for many high-throughput applications.\end{tabular} \\ \hline
\textbf{Energy Efficiency} &
  \begin{tabular}[c]{@{}l@{}}High, optimized for power-sensitive \\ environments.\end{tabular} &
  \begin{tabular}[c]{@{}l@{}}Moderate to high, depends on workload \\ and optimization techniques.\end{tabular} \\ \hline
\textbf{Programmability} &
  \begin{tabular}[c]{@{}l@{}}Difficult, requires expertise in hardware \\ description languages (HDLs).\end{tabular} &
  \begin{tabular}[c]{@{}l@{}}Easier, supported by mature software \\ ecosystems (CUDA, OpenCL).\end{tabular} \\ \hline
\textbf{Flexibility} &
  \begin{tabular}[c]{@{}l@{}}High, can be reconfigured for specific \\ tasks.\end{tabular} &
  \begin{tabular}[c]{@{}l@{}}Low, fixed architecture with limited \\ customization options.\end{tabular} \\ \hline
\textbf{Development Cost} &
  \begin{tabular}[c]{@{}l@{}}High initial effort due to complex design \\ and validation processes.\end{tabular} &
  \begin{tabular}[c]{@{}l@{}}Lower compared to FPGAs, faster \\ prototyping with available frameworks.\end{tabular} \\ \hline
\end{tabular}%
\end{table*}

\begin{table*}[]
\centering
\caption{Domain-specific information of reviewed papers and the performance metrics.
}
\label{domain}
\begin{tabular}{|c|c|c|c|c|c|}
\hline
\textbf{\begin{tabular}[c]{@{}c@{}}Application\\ Category\end{tabular}} &
  \textbf{Throughput} &
  \textbf{Latency} &
  \textbf{\begin{tabular}[c]{@{}c@{}}Energy \\ efficiency\end{tabular}} &
  \textbf{Programmability} &
  \textbf{Papers} \\ \hline
\textbf{\begin{tabular}[c]{@{}c@{}}Massively Parallel\\ Applications\end{tabular}}         & \checkmark & \checkmark &  \checkmark &  \checkmark &  \cite{jason, elsev, socccloud,soccopencl,socctaxonomy,hpec,sparse}\\ \hline
\textbf{\begin{tabular}[c]{@{}c@{}}AI and Deep \\ Learning\end{tabular}}                   & \checkmark & \checkmark &  \checkmark &  \checkmark&  \cite{aicomp, aiic, dnn} \\ \hline
\textbf{\begin{tabular}[c]{@{}c@{}}Vision Processing and \\ Embedded Systems\end{tabular}} & \checkmark & \checkmark &  \checkmark & - &   \cite{vision} \\\hline
\end{tabular}%

\end{table*}

\section{Methodology}
\label{sec:method}

This study compiles and synthesizes insights from multiple research articles comparing FPGA and GPU performance across various benchmarks and application domains. The methodology followed a systematic approach consisting of the following steps:

\begin{enumerate}

\item Study Selection and Benchmarking:
We reviewed a diverse set of research articles and papers that provided quantitative comparisons between FPGAs and GPUs. The selection criteria included: i. papers that present benchmarks from well-established libraries and frameworks \cite{dnn}. ii. Studies covering a broad range of application domains ranging from deep learning, graph transversal, structured grids, data mining, image processing, physics simulation, linear algebra, etc. \cite{rodinia, soccopencl} In terms of performance metrics, papers reporting execution time, energy efficiency, and memory bandwidth utilization were covered on both FPGA and GPU platforms.

\item Categorization of application profiles We classified the workloads discussed in the reviewed papers based on their computational and memory characteristics:
\emph{Regular vs Irregular Workloads:} Workloads were classified as regular if they exhibited predictable control flow and memory access patterns, such as structured grids and image processing tasks. Irregular workloads, such as graph traversal and sparse matrix operations, involve dynamic control flow and scattered memory access patterns.

\emph{Memory-Intensive vs Compute-Intensive Workloads:} We further distinguished workloads based on their dependency on memory bandwidth and compute throughput. Memory-intensive applications often experience bottlenecks due to memory stalls, while compute-intensive applications are limited by raw processing power.

\item Compilation of Performance Metrics
We compiled and compared results from the reviewed studies using the following key metrics:

\emph{Execution Time:} To evaluate raw performance across FPGA and GPU platforms.

\emph{Energy Efficiency:} To assess the power consumption relative to performance, particularly important in embedded and real-time systems.

\emph{Memory Bandwidth Utilization:} To understand how efficiently each accelerator handles data-intensive workloads.

\end{enumerate}

Table \ref{domain} lists the types of papers reviewed for this article. By systematically categorizing workloads and compiling performance data from various studies, this survey aims to provide actionable guidance for hardware selection based on specific application profiles.

\vspace{0.3cm}
\section{Key insights and recommendations}
\label{sec:insights}

The selection of the right hardware accelerator, FPGA or GPU, is highly dependent on the characteristics of the workload. While GPUs are widely known for their ease of development and high throughput in parallelizable tasks, FPGAs offer unmatched flexibility and energy efficiency for custom domain-specific tasks. A detailed understanding of the dependencies and computational patterns of an application (parallelizable kernel code) enables informed decision-making when selecting between these accelerators. This section provides a practical framework for guiding hardware selection based on two key factors:- \emph{Dependency Analysis using LLVM} and \emph{Application profile}

\begin{table*}[h]
\centering
\caption{FPGA vs GPU Performance: Dependency-Based Comparison}
\label{perftab}
\resizebox{\textwidth}{!}{%
\begin{tabular}{|l|c|c|c|c|}
\hline
\multicolumn{1}{|c|}{\textbf{Application}} &
  \textbf{\begin{tabular}[c]{@{}c@{}}Percentage of \\ decouplable \\ variables(\%)\end{tabular}} &
  \textbf{\begin{tabular}[c]{@{}c@{}}Performance comparison \\ (Execution time)\\ Intel FPGA\end{tabular}} &
  \textbf{\begin{tabular}[c]{@{}c@{}}Performance comparison  \\ (Execution time)\\ Xilinx FPGA\end{tabular}} &
  \textbf{\begin{tabular}[c]{@{}c@{}}Performance/Watt\\ Intel FPGA\end{tabular}} \\ \hline
\textbf{Nearest Neighbor} & 100 & Comparable & Comparable & FPGA wins \\ \hline
\textbf{Srad}             & 100 & Comparable & Comparable & FPGA wins \\ \hline
\textbf{B+Tree}           & 60  & GPU wins   & X          & GPU wins  \\ \hline
\textbf{BFS}              & 40  & GPU wins   & GPU wins   & GPU wins  \\ \hline
\textbf{LUD Diag}         & 100 & FPGA wins  & FPGA wins  & X         \\ \hline
\textbf{Hotspot}          & 100 & GPU wins   & GPU wins   & X         \\ \hline
\end{tabular}%
}
\end{table*}

\subsection {Dependency Analysis using LLVM}

Modern workloads often involve complex dependencies between memory and computation, which significantly impact performance on hardware accelerators. Low-Level Virtual Machine \cite{llvm,llvm1} is a widely used compiler framework that provides tools for static analysis and code optimization. In \cite{elsev}, an LLVM-based dependency analysis tool was developed to identify decouplable variables and detect loop-carried dependencies in the application code. The LLVM tool statically analyzes the kernel code to detect dependencies between operations and memory accesses. Specifically, it identifies:

\begin{itemize}
\item Decouplable variables: 
\end{itemize}
Variables whose memory access patterns can be isolated from the computation logic, meaning memory operations can be performed independently of the core computation. Identifying such variables is crucial because it allows memory operations to overlap with computation, reducing idle time, and improving overall throughput. In FPGA-based systems, this enables efficient pipelining, where memory fetches and computations occur in parallel, minimizing latency. On GPUs, decouplable variables benefit from the ability to hide memory latency by keeping more threads active while waiting for memory operations to complete. 

\begin{itemize}
\item Loop-carried dependencies: 
\end{itemize}
Loop-carried dependencies occur when iterations of a loop are interdependent—i.e., an iteration depends on the results of the previous iteration. Such dependencies limit the ability to parallelize the loop, as iterations cannot be executed simultaneously without violating data correctness. On FPGAs, loop-carried dependencies can reduce the effectiveness of pipelining as the execution of each iteration must wait for the completion of the previous one. On GPUs, while loop-carried dependencies can also limit parallelism, the hardware’s dynamic scheduling capabilities and large number of threads often mitigate this issue by executing independent iterations concurrently whenever possible. 

To gain deeper insights into the findings, several related works \cite{elsev,socccloud,soccopencl,hpec,socctaxonomy} were compiled (Table \ref{perftab}). All these works utilized the Rodinia benchmark suite \cite{rodinia} deployed across different hardware platforms (Table \ref{accspec}). The study presented in \cite{elsev} was taken as the baseline, and comparisons were made with other works based on identical kernel code for the selected applications. This ensured that the performance variations observed were primarily due to hardware differences rather than differences in implementation.

The results were compiled into two key metrics: raw performance (measured as execution time) and performance per watt (energy efficiency). Exact workload sizes were maintained across all applications to ensure a fair comparison. For each application, results are highlighted as follows:

\begin{itemize}
    \item GPU wins:  If the GPU outperforms the FPGA in both execution time and energy efficiency.
    \item FPGA wins: If the FPGA showed better or comparable performance in terms of execution time and energy efficiency.
    \item Comparable: If the performance metrics were similar for both platforms.
    \item X: If data for a specific application on a given platform was unavailable.
\end{itemize}

\begin{table}[]
\centering
\caption{Accelerator hardware specifications}
\label{accspec}
\begin{tabular}{|l|c|c|}
\hline
\multicolumn{1}{|c|}{\textbf{\begin{tabular}[c]{@{}c@{}}Accelerator\\ Type\end{tabular}}} &
  \textbf{\begin{tabular}[c]{@{}c@{}}Device \\ Family\end{tabular}} &
  \textbf{Specifications} \\ \hline
\textbf{Intel FPGA} & Stratix V                                               & \begin{tabular}[c]{@{}c@{}}CLBs 234,700\\ Registers 939,000\\ DSP  blocks 256\end{tabular} \\ \hline
\textbf{Xilinx FPGA} &
  \begin{tabular}[c]{@{}c@{}}Virtex\\ Ultrascale \\ VU9P\end{tabular} &
  \begin{tabular}[c]{@{}c@{}}CLBs 2,364,000\\ Registers 2,330,479\\ DSP blocks 1248\end{tabular} \\ \hline
\textbf{AMD GPU}    & \begin{tabular}[c]{@{}c@{}}Firepro\\ W7100\end{tabular} & GPU max compute uinits 28                                                                  \\ \hline
\end{tabular}%
\end{table}

The selected applications represent a broad range of computational patterns and domains, including structured grids, graph traversal, data mining, image processing, physics simulation, and linear algebra. The results indicate that for applications with a high percentage of decouplable variables, FPGAs exhibited superior or comparable performance relative to GPUs. This supports the hypothesis that FPGAs excel in workloads where dependencies can be decoupled and parallelism can be exploited through customized pipelines.

On the other hand, a thorough comparison of applications with a high proportion of non-decouplable variables was not feasible due to limited data. In the two cases where sufficient data was available—B+Tree and BFS (Breadth-First Search)—GPUs outperformed Xilinx FPGAs, while comparable performance was observed on Intel FPGAs. This result aligns with the hypothesis that GPUs are better suited for workloads with high loop-carried dependencies and irregular memory access patterns, where their massive thread-level parallelism and dynamic scheduling capabilities can be fully leveraged.

Interestingly, for Hotspot, an application with 100\% decouplable variables, GPUs still performed better. This suggests that while dependency analysis is an important factor, the nature of the application—such as its computation-to-memory ratio and parallelism granularity also play a crucial role in determining which hardware accelerator yields superior performance.


\subsection{Application profile}

Application profiles are defined by characteristics such as regularity of operations, memory access patterns, and sensitivity to latency and bandwidth. Regular applications exhibit predictable control flow and structured memory access, whereas irregular applications involve dynamic control flow and non-uniform memory access, which can complicate parallel execution. Additionally, understanding how memory stalls and bandwidth utilization affect performance is crucial for evaluating the suitability of FPGAs and GPUs in handling memory-intensive workloads. 


\vspace{5pt}
\begin{itemize}
    \item Stalls and Bandwidth:
\end{itemize}

Effective memory management is critical for hardware accelerator performance, with FPGAs and GPUs adopting different strategies. GPUs use built-in schedulers and memory hierarchies to hide latency by switching between active threads during memory stalls. In contrast, FPGAs achieve efficiency through customized data paths, fine-grained parallelism, and deep pipelining. However, without a runtime scheduler or multi-level cache, FPGAs struggle with memory-intensive workloads, where stalls can cause severe data-path underutilization and bandwidth inefficiency.

\begin{figure}[h]
	\vspace{-5pt}
		\centering
				\subfloat[Memory stalls]{\includegraphics[width=.5\linewidth]{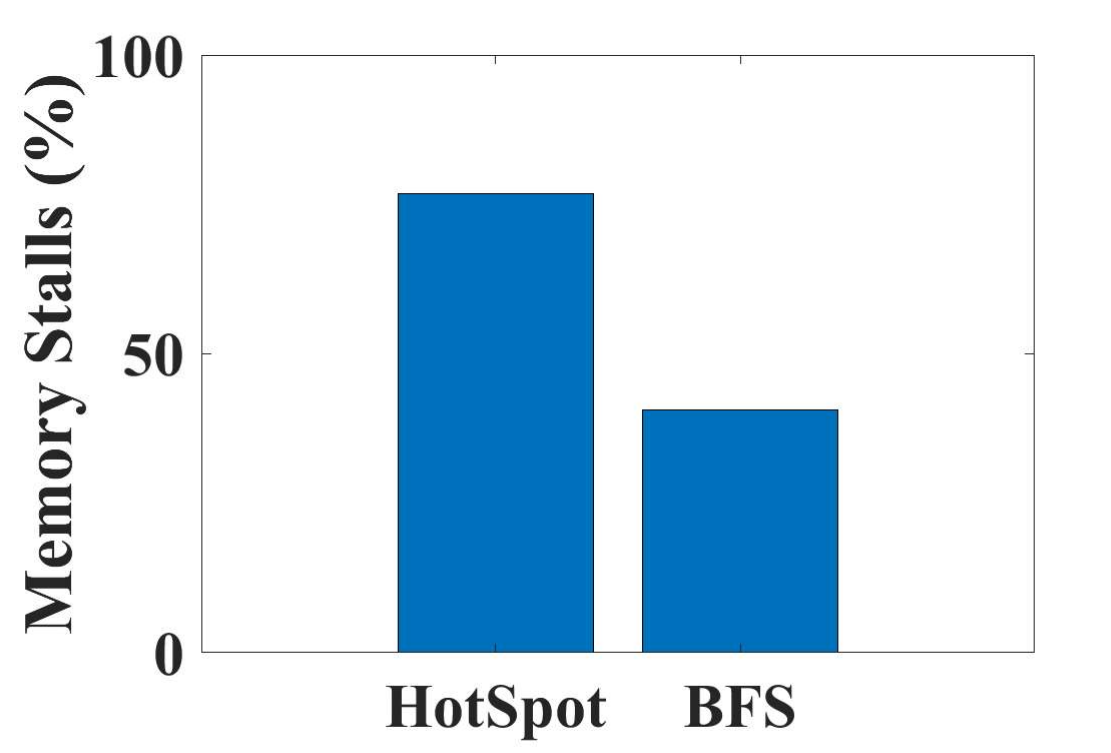}\label{fig:stall_intro}}
				\hfill
				\subfloat[Memory bandwidth utilization]{\includegraphics[width=.5\linewidth]{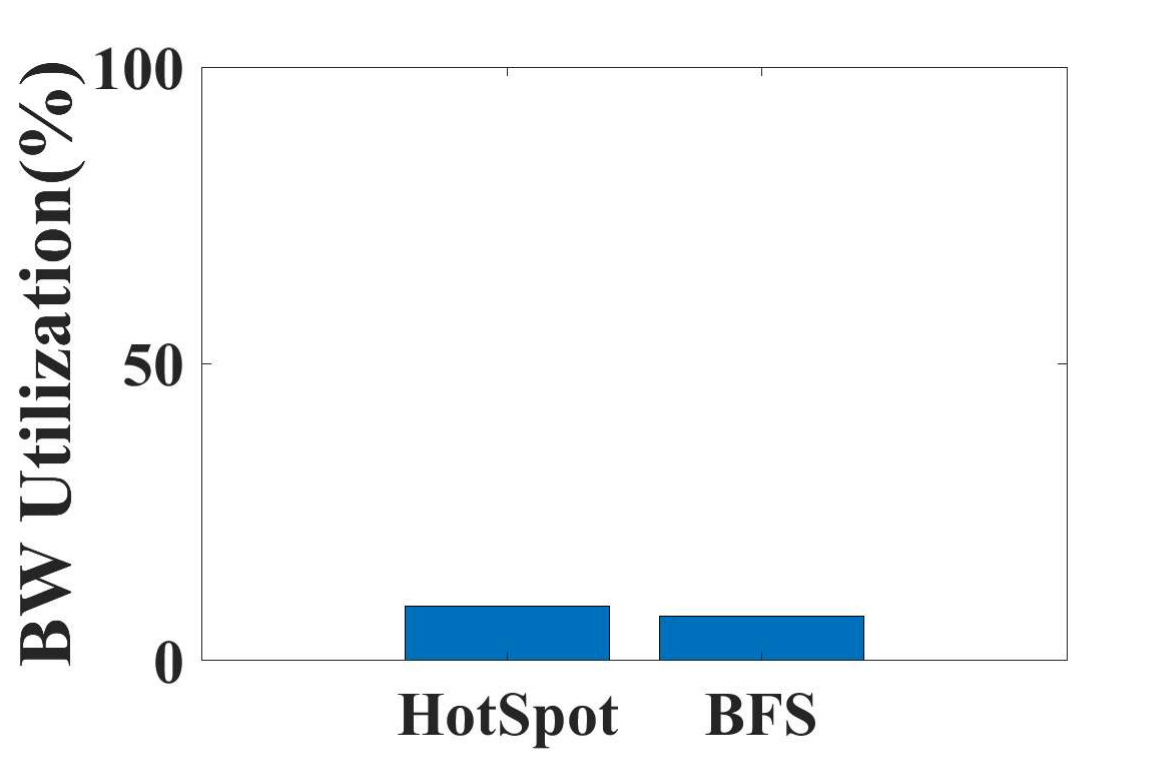}\label{fig:bw_intro}}
									\vspace{-5pt}
				\caption{Memory stalls and bandwidth utilization for Hotspot and BFS applications}
				\label{fig:intro}
					\vspace{-5pt}
\end{figure}

This problem has been extensively studied in \cite{elsev}. Since Hotspot and BFS were analyzed in the previous section, we continue using these two memory-intensive applications from the Rodinia benchmark suite (Fig. \ref{fig:stall_intro} and \ref{fig:bw_intro}) as illustrative examples. In these applications, 75\% and 40\% of memory accesses, respectively, result in stalls within the pipeline. As a result, memory bandwidth utilization drops to just 9\% in Hotspot and 7\% in BFS, severely hindering performance. In such cases a GPU is bound to perform better than an FPGA. These findings underscore the importance of understanding an application’s memory access patterns when selecting between FPGA and GPU accelerators.

\vspace{5pt}
\begin{itemize}
    \item Memory Access Patterns:
\end{itemize}

Memory access patterns refer to how applications retrieve data from memory—either in a predictable, structured manner (regular) or in a more dynamic, data-dependent manner (irregular). Regular memory access patterns are sequential or follow a known structure, making them easier to optimize through techniques such as pipelining and efficient prefetching. Irregular patterns, on the other hand, involve unpredictable or scattered memory accesses, often dependent on runtime data, making optimization more challenging.

In terms of accelerator suitability, FPGAs tend to excel in applications with regular memory access patterns, where their ability to implement custom pipelines and efficient memory controllers can be fully utilized \cite{dnn}. In contrast, GPUs are better equipped to handle irregular memory access patterns due to their high thread count and dynamic scheduling capabilities, which allow them to hide memory latency by switching between active threads \cite{aicomp,aiic}.

In our chosen examples, the BFS application exhibits irregular memory access due to its graph traversal nature. Thus GPUs demonstrate superior performance in this case. Conversely, Hotspot, with its regular grid-based memory access pattern, is better suited for FPGAs. However, due to the large number of memory stalls (as seen in the previous section), this doesn't turn out to be the case. Similarly, in deep learning models with dense computations and regular memory patterns (e.g., convolutional neural networks in vision tasks), FPGAs can be effective \cite{dnn, vision}, while GPUs excel in sparse matrix operations \cite{sparse} or models requiring irregular data access \cite{soccopencl,socccloud}.

\vspace{5pt}
\begin{itemize}
    \item Latency and Execution Model:
\end{itemize}

Latency refers to the time required for an accelerator to process a single task or operation from start to finish. In real-time and latency-sensitive applications, such as embedded vision systems, signal processing, and edge AI inference, minimizing latency is crucial for maintaining responsiveness. Beyond latency itself, the execution model—how tasks are scheduled and executed—plays a significant role in determining performance, particularly for parallel workloads.

FPGAs excel in low-latency environments due to their deterministic execution model. By implementing custom hardware pipelines tailored to specific applications, FPGAs can achieve highly predictable, low-latency processing. Additionally, since FPGAs lack a global scheduler, there is minimal overhead, enabling faster task completion compared to GPUs in latency-critical scenarios.

GPUs, on the other hand, employ a massively parallel execution model, designed to maximize throughput rather than minimize latency. Their high latency often arises from kernel launch overhead, thread scheduling, and memory hierarchy management. While this can be a disadvantage in latency-sensitive applications, GPUs are optimized for throughput-oriented tasks where large batches of data can be processed concurrently, such as deep learning model training and large-scale data processing.

In summary, FPGAs are ideal for applications requiring low latency and deterministic execution, while GPUs are better suited for high-throughput workloads where latency is less critical. 

\subsection{Final thoughts- Guidance for Accelerator Selection} 

Selecting the right hardware accelerator—FPGA or GPU—depends on several key factors, including workload dependencies, memory access patterns, and latency requirements. By combining insights from dependency analysis, application profiling, and execution models, users can make informed decisions tailored to their specific workloads. Below is a summary of when each accelerator may be more suitable:

\vspace{10pt}
\begin{itemize}
    \item FPGA Preference:
\end{itemize}

FPGAs are generally preferred for workloads where the LLVM-based tool \cite{elsev} identifies a high percentage of decouplable variables and predictable memory access patterns. These characteristics allow FPGAs to efficiently pipeline memory operations and computation, minimizing memory stalls and delivering low-latency performance. FPGAs excel in tasks requiring fine-grained control and deterministic execution, such as real-time embedded systems, signal processing, and low-latency financial applications. Additionally, applications with regular memory access patterns, such as structured grids and certain dense deep learning models, benefit from FPGA’s custom data paths and low-latency execution.

\vspace{5pt}
\begin{itemize}
    \item GPU Preference:
\end{itemize}

GPUs are better suited for workloads with a high number of loop-carried dependencies, irregular memory access patterns, or tasks demanding high throughput. The dynamic scheduling capabilities and high memory bandwidth of GPUs make them ideal for handling irregular applications, such as graph traversal, sparse matrix operations, and AI model training. Additionally, GPUs excel in scenarios where latency can be traded off for high throughput, such as large-scale data processing and batch-oriented tasks.

\vspace{0.3cm}
\section{Challenges and Future Directions}
\label{sec:challenges}

Despite significant advancements, FPGAs and GPUs still have limitations that affect informed decision-making. For FPGAs, programmability remains a key challenge, despite the availability of improved HLS toolchains and broader OpenCL support. Additionally, while newer FPGAs feature high-bandwidth memory (HBM), fully utilizing this bandwidth demands careful memory access optimization. GPUs, meanwhile, remain less power-efficient compared to FPGAs.

Future work could focus on expanding the LLVM-based toolchain to include application profiles and workload characteristics, creating a complete decision-support framework. Such a toolchain would allow developers to input application code and receive recommendations on hardware selection. Additionally, research into hybrid architectures combining the strengths of FPGAs and GPUs could enable optimal performance across a broader range of applications.
\vspace{0.3cm}
\section{Conclusion}
\label{sec:Conclusion}

This paper surveyed various studies comparing FPGAs and GPUs, providing insights into hardware selection based on dependency analysis, memory access patterns, and application profiles. While FPGAs excel in workloads requiring low latency and customized pipelines, GPUs are better suited for high-throughput tasks with irregular memory access. Despite advancements in programmability, memory bandwidth for FPGAs and power efficiency for GPUs, challenges remain. Future work could focus on creating comprehensive tool-chains that automate hardware recommendations, aiding developers in optimizing performance across a wide range of applications.

\vspace{0.3cm}

\bibliographystyle{IEEEtran}
\scriptsize
\bibliography{ieee}

\end{document}